\documentclass[aps,twocolumn,prl,  nofootinbib]{revtex4}

\usepackage[mathscr]{eucal}
\usepackage{amsfonts,amssymb,amsmath,bbm}

\usepackage[all]{xy}  
\usepackage{tikz}

\newcommand\beq{\begin{equation}}
\newcommand\eeq{\end{equation}}
\newcommand\be{\begin{eqnarray}}
\newcommand\ee{\end{eqnarray}}
\newcommand\beqa{\begin{eqnarray}}
\newcommand\eeqa{\end{eqnarray}}
\newcommand\bean{\begin{eqnarray*}}
\newcommand\eean{\end{eqnarray*}}
\newcommand{\bes}{\begin{eqnarray}}
\newcommand{\ees}{\end{eqnarray}}

\newcommand{\cO}{{\mathcal O}}

\newcommand{\cS}{{\mathcal S}}

\newcommand\N{{\mathbb N}}
\newcommand\R{{\mathbb R}}
\newcommand\Z{{\mathbb Z}}

\newcommand{\SO}{{\rm SO}}
\newcommand{\so}{\mathfrak{so}}

\newcommand{\su}{\mathfrak{su}}
\newcommand{\SU}{{\rm SU}}
\newcommand{\g}{{\mathfrak{g}}}

\def\inv{{\mbox{\tiny -1}}}
\def\minus{{\mbox{\small -}}}
\def\plus{{\mbox{\tiny +}}}

\newcounter{letter} \newcounter{numeral} \newcounter{Numeral}

\newcommand\Tr{\mathrm{Tr}}
\def\extd{\mathrm {d}}
\def\vphi{\varphi}
\def\vphihat{\widehat{\varphi}}
\newcommand\e{{\mbox{e}}}

\begin{document}

\title{
Group field theory with non-commutative metric variables\\
}

\author{{\bf Aristide Baratin} and {\bf Daniele Oriti}}
\affiliation{Albert Einstein Institute, Am M\"uhlenberg 1, D-14476 Golm, Germany \\ aristide.baratin@aei.mpg.de \hspace{1cm} daniele.oriti@aei.mpg.de}

\begin{abstract}
We introduce a dual formulation of group field theories, making them a type of non-commutative field theories. In this formulation, the variables of the field are Lie algebra variables with a clear interpretation in terms of simplicial geometry. For Ooguri-type models, the Feynman amplitudes are simplicial path integrals for $BF$ theories.  This formulation suggests ways to impose the simplicity constraints involved in $BF$ formulations of 4d gravity directly at the level of the group field theory action. We illustrate this by giving a new GFT definition of the Barrett-Crane model. 
\end{abstract}

\maketitle
\noindent Group field theories \cite{gft} (GFTs) are developing into a promising formalism for quantum gravity, combining elements from several approaches \cite{libro}. They are a higher-dimensional generalization of matrix models for 2d gravity  \cite{mm} and build up on the achievements of loop quantum gravity \cite{lqg}  and spin foam models \cite{SF}. Loop quantum gravity describes quantum space in terms of spin networks; spin foam models define its dynamics in a covariant language. GFTs subsume this dynamics, as every spin foam model can be interpreted as a GFT Feynman amplitude. 

Several results in spin foam models, and the historic roots in matrix models, suggest a close relation between GFTs and simplicial gravity path integrals, as used in other discrete approaches \cite{simplicial}. The spin foam quantization is based on the geometric quantization of simplicial geometry \cite{SF,new}, and there are close relations between simplicial and loop gravity canonical data \cite{biancajimmy}. 
The resulting amplitudes contain the Regge action in a semi-classical limit \cite{asymp}, and can be analyzed from a path integral perspective \cite{valentin}. However much remains to be understood, and GFTs, which already realize a duality between spin foam models and lattice path integrals in connection variables, 
seem a convenient setting to do so.

In parallel, interesting connections between spin foam/GFT models and  non-commutative geometry have been discovered. 
Non-commutative matter field theories, interesting for quantum gravity phenomenology \cite{QGphen}, can be derived either from coupling particles to spin foam amplitudes \cite{PR3}, or from GFT actions, as perturbations around classical GFT solutions \cite{eterawinston}. 
These results suggest that non-commutative structures lie hidden at the very foundations of the GFT formalism.   

In this paper we realize the above suggestions. We recast GFTs as non-local, non-commutative field theories on Lie algebras, which we relate to the B variables of simplicial BF theory. We prove that the Feynman amplitudes for arbitrary diagram are simplicial BF path integrals. This new representation of GFTs gives an explicit duality between spin foam models and simplicial gravity path integrals,  
and clarifies the encoding of  simplicial geometry in the action. 
We illustrate this by giving a new GFT definition of the Barrett-Crane model. 
We leave the details of our construction to a following publication. 

\noindent {\bf Non-commutative representation of 3d GFT.} 
We first consider Boulatov's group field formulation of 3d Riemannian gravity \cite{Boulatov}. The variables are fields $\varphi_{123} \!:=\!\varphi(g_1, g_2, g_3)$ on $\SO(3)^3$ satisfying the invariance:
\beq \label{invariance}
\vphi(g_1, g_2, g_3)  =  \vphi(hg_1, hg_2, hg_3) 
\eeq
$\forall h \in \SO(3)$. The dynamics is governed by the action: 
\be
S =\! \frac{1}{2}\int [\extd g]^3 \,\varphi^2_{123} -  \frac{\lambda}{4!} \! \int [\extd g]^6 \,\varphi_{123}\varphi_{345}\varphi_{526}\varphi_{641} \label{boulatov}
\ee
The Feynman graphs generated by this theory are 2-complexes dual to 3d triangulations: the combinatorics of the field arguments in the interaction vertex is that of a tetrahedron, while the kinetic term dictates the gluing rule for tetrahedra along triangles. 
The Peter-Weyl theorem gives an expansion of $\varphi$ in terms of functions on irreducible representations  $j_i \in \N$ of $\SO(3)$. In such a representation,  the field is pictured as 3-valent spin network vertex and interpreted as a quantized triangle;  
a generic Feynman amplitude gives the well-known Ponzano-Regge spin foam model \cite{Boulatov}. 



We now introduce an alternative formulation of the model, 
obtained by a `group Fourier transform' \cite{PR3, NCFourier} mapping  functions on a  group $G$ to (non-commutative) functions on its Lie algebra $\g$. The transform is based on plane waves $\e_{g}(x) \!=\! e^{i \vec{p}_g\cdot \vec{x}}$, labelled by $g\in G$,  as functions on $\g \sim \R^n$, depending on a choice of coordinates $\vec{p}_g$ on the group manifold. In what follows we will identify functions of $\SO(3)$ with functions of $\SU(2)$ invariant under $g \to - g$. 

We choose  the coordinates $\vec{p}_g = \Tr (|g| \vec{\tau})$, where $|g| \! := \! \mbox{sign} (\Tr g)g$, $\vec{\tau}$ are $i$ times the Pauli matrices and `$\Tr$' is the trace in the fundamental representation. 
For $x \!=\! \vec{x} \cdot \vec{\tau}$ and 
$g \!=\! e^{\theta \vec{n} \cdot \vec{\tau}}$, we thus have 
\beq
\e_g(x)  =  e^{i \Tr (x |g|)} = e^{-2 i\sin \theta \vec{n} \cdot \vec{x}}
\eeq
The Fourier transform of functions $f$ on $\SU(2)$ is defined by
\beq
\widehat{f}(x) = \int \extd g f(g)\, \e_g(x)
\eeq
where $\extd g$ is the normalized Haar measure. 

The image of the Fourier transform inherits an algebra structure from the convolution product on the group, given by the $\star$-product defined on plane waves as
\beq
\e_{g_1}\star \e_{g_2} = \e_{g_1g_2} 
\eeq
On functions of $\SO(3)$, the Fourier is invertible:
\beq
f(g) =  \frac{1}{\pi}\int \extd^3 x \, (\widehat{f} \star \e_{g^\inv})(x)
\eeq
With a bit more work  the above construction is extended to an invertible $\SU(2)$ Fourier transform \cite{NCFourier}.

Fourier transform and $\star$-product extend to functions of several variables like the Boulatov field as
\beq
\vphihat_{123}\, :=\, \hat{\varphi}(x_1, x_2, x_3) =\int [\extd g]^3\, \vphi_{123} \, \e_{g_1}(x_1) \e_{g_2}(x_2) \e_{g_3}(x_3) 
\eeq
The first feature of the dual formulation is that the constraint (\ref{invariance}) acts on dual fields as a `closure constraint' for the variables $x_j$.
Indeed, given the projector $P\vphi_{123} \!:=\! \int \extd h \, \vphi(hg_1, hg_2, hg_3)$ onto gauge invariant fields, 
a simple calculation gives 
\beq
\widehat{P\vphi} = \widehat{C} \star \vphihat, \quad \widehat{C}(x_1, x_2, x_3) = \delta_0(x_1 \!+\!x_2\!+\!x_3) 
\eeq
where $\delta_0$ is the element $x=0$ of the family of functions:
\beq \label{delta}
\delta_x(y)  := \frac{1}{\pi}\int_{\SO(3)} \extd h \, \e_{h^\inv}(x)  \e_h(y)
\eeq
These play the role of Dirac distributions in the non-commutative setting, in the sense that
\beq \label{Dirac}
\int \extd^3 y \, (\delta_x \star f)(y) = \int \extd^3 y\, (f \star \delta_x)(y) = f(x) 
\eeq
We may thus interpret the variables of the Boulatov dual field as the edges vectors of a triangle in $\R^3$, and the dual fields themselves as (non-commutative) triangles.

Since the $\star$-product is dual to group convolution, the combinatorial structure of the action in terms of the dual field matches the one in (\ref{boulatov}). We may thus show that
\beq
S =\! \frac{1}{2}\int [\extd x]^3 \,\vphihat_{123} \star \vphihat_{123} -  
\frac{\lambda}{4!} \! \int [\extd x]^6 \,\vphihat_{123} \star \vphihat_{345} \star \vphihat_{526} \star \vphihat_{641}
\eeq
where it is understood that $\star$-products relate repeated indices as 
$\phi_i \star \phi_i \! :=\! (\phi \star \phi_{\minus})(x_i)$, with $\phi_{\minus}(x) \! := \! \phi(\minus x)$.
The structure of this action is best visualized in terms of diagrams. Thus,  kinetic and interaction terms identify a propagator and a vertex given by:
\vspace{-1mm}
\beq \label{Feynmanrules}
\begin{array}{c}
\begin{tikzpicture}[scale=1.2]
\draw (-0.3,-0.2) rectangle (0.3, 0.2);
\path (0, 0) node {\small $t$};
\draw (-0.2, -1) -- (-0.2,-0.2);
\draw (0,-1) -- (0, -0.2);
\draw (0.2, -1) -- (0.2, -0.2);
\path (-0.2, -1.2) node {\tiny $x_1$};
\path (0, -1.2) node  {\tiny $x_2$};
\path (0.2, -1.2) node {\tiny $x_3$};
\path (0, -1.5) node  {\tiny $ $};

\draw (-0.2, 0.2) -- (-0.2, 1);
\draw (0, 0.2) -- (0, 1);
\draw (0.2, 0.2) -- (0.2, 1);
\path (0.2, 1.2) node {\tiny $y_3$};
\path (0, 1.2) node  {\tiny $y_2$};
\path (-0.2, 1.2) node {\tiny $y_1$};
\path (0, 1.5) node  {\tiny $ $};

\end{tikzpicture}
\hspace{1cm} 
\begin{tikzpicture}[scale=1.2]
\draw (-1, 0) -- (1,0);
\draw (0,-1) -- (0, -0.1);
\draw (0, 0.2) -- (0, 1);
\draw (-1, 0.2) -- (-0.2, 0.2);
\draw (0.2, 0.2) -- (1, 0.2);
\draw (-1, -0.2) -- (-0.2, -0.2);
\draw (0.2,- 0.2) -- (1, -0.2);
\draw (-0.2, -1) -- (-0.2,-0.2);
\draw (-0.2, 0.2) -- (-0.2, 1);
\draw (0.2, -1) -- (0.2, -0.2);
\draw (0.2, 0.2) -- (0.2, 1); 
\path (0, 0.1) node {\small $\tau$};
\path (-1.2, 0.2) node {\tiny $x_1$};
\path (-1.2, 0) node {\tiny $x_2$};
\path (-1.2, -0.2) node {\tiny $x_3$};
\path (-0.2, -1.2) node {\tiny $y_3$};
\path (0, -1.2) node  {\tiny $x_4$};
\path (0.2, -1.2) node {\tiny $x_5$};
\path (1.2, -0.2) node {\tiny $y_5$};
\path (1.2, 0) node {\tiny $y_2$};
\path (1.2, 0.2) node {\tiny $x_6$};
\path (0.2, 1.2) node {\tiny $y_6$};
\path (0, 1.2) node  {\tiny $y_4$};
\path (-0.2, 1.2) node {\tiny $y_1$};
\path (-1.5, 0) node {\tiny $t_a$};
\path (0, -1.4) node  {\tiny $t_b$};
\path (1.5, 0) node {\tiny $t_c$};
\path (0, 1.4) node  {\tiny $t_d$};
\end{tikzpicture}
\end{array}
\eeq
\vspace{-8mm}
\beq  \label{prop&vertex}
\int \extd h_t \, \prod_{i=1}^3 (\delta_{\minus x_i} \star \, \e_{h_t} )(y_i), 
\quad 
\int \prod_t \extd h_{t\tau} \, \prod_{i=1}^6 ( \delta_{\minus x_i} \star\, \e_{h_{tt'}})(y_i)
\eeq
with $h_{tt'} := h_{t\tau} h_{\tau t'}$. We have used `$t$' for triangle and `$\tau$' for tetrahedron. The group variables $h_{t}$ and $h_{t\tau}$ arise from gauge invariance  (\ref{invariance}). 

The integrands in (\ref{prop&vertex}) factorize into a product of functions associated to strands (one for each field argument), with a clear geometrical meaning. Just like in the standard group representation \cite{gft},  the group elements $h_{t}$ and $h_{t\tau}$  are interpreted as parallel transports through the triangle $t$, and from the  center of the tetrahedron $\tau$ to triangle $t$, respectively. 
The pair of variables $(x_i,y_i)$ associated to the same edge $i$ corresponds to the edges vectors seen from the frames associated to the two triangles $t, t'$ sharing it. The vertex functions state that the two variables are identified, up to parallel transport $h_{tt'}$, and up to a sign labeling the two opposite edge orientations inherited by the  triangles $t, t'$. The propagator encodes a similar gluing condition, allowing for the possibility of a further mismatch between the reference frames associated to the same triangle in different tetrahedra.


{\bf Feynman amplitudes as simplicial path integrals.}
\noindent 
In building up a closed Feynman graph, propagator and vertex strands are joined to one another using the $\star$-product, keeping track of the ordering of functions associated to the various building blocks of the graph. Each loop of strands bound a face of the 2-complex, which is dual to an edge of the triangulation. 

Under the integration over the group variables $h_t, h_{t\tau}$,  the amplitude factorizes into a product of {\it face} amplitudes. 
Let $f_e$ be a face of the 2-complex, dual an edge $e$ in the triangulation, and consider the loop of strands that bound it. 
The choice of an orientation and a reference vertex defines an ordered sequence $\{\tau_j\}_{0\leq j\leq N}$ of vertices on the loop (equivalently, an ordered set of tetrahedra around $e$). Using (\ref{Dirac}), each vertex $\tau_j$, after contraction with the propagator $t_j$ joining $\tau_j$ and $\tau_{j\plus1}$, contributes with $(\delta_{x_j} \star \e_{h_{jj\plus1}})(x_{j\plus1})$ to the face amplitude, where $h_{jj\plus1} \!=\! h_{\tau_j t_j} h_{t_j} h_{t_j\tau_{j\plus 1}}$ parallel transports $j$ to $j+1$. 

The face amplitude $A_{f_e}[h]$ is then the cyclic $\star$-product of all these contributions: 
\beq
A_{f_e}[h] = \int \prod_{j=0}^N \extd^3 x_j \vec{\bigstar}_{j=0}^{N + 1} \, (\delta_{x_j} \star \e_{h_{jj+1}})(x_{j+1}) 
\eeq
where $x_{N+1} \!:=\! x_0$. This amplitude encodes the identification, up to parallel transport, of the metric variables associated to $e$ in different tetrahedron frames. 
Integrating over all metric variables $x_j$ in $A_{f_e}[h]$, except for that of the reference frame, we obtain the Feynman amplitude: 
\beq \label{bf} 
Z(\Gamma) = \frac{1}{\pi^{|e|}}\int \prod_t \extd h_t \prod_e \extd^3 x_e  \, e^{i \sum_e \Tr \, x_e H_e}
\eeq
where $h_t$ is the parallel transport between the two tetrahedra sharing $t$ and $H_e$ is the holonomy around the boundary of $f_e$, computed from a given tetrahedron, and $|e|$ is the number of edges of the triangulation. 

Equ.(\ref{bf}) is the usual expression for the simplicial path integral of first order 3d gravity. The Lie algebra variables $x_e$, one per edge of the simplicial complex, play the role of discrete triad;  the group elements $h_t$, one per triangle or link of the dual 2-complex, play the role of discrete connection, defining the discrete curvature $H_e$ through holonomy around the faces dual to the edges of the simplicial complex. 

Open GFT Feynman amplitudes have fixed boundary simplicial data. 
The one-vertex contribution to the 4-point functions, for example, is the function of twelve metric variables $x_i, x'_i$ obtained by acting with a closure operator $\widehat{C}$ (propagator) on each external 3-stranded leg of the vertex diagram in (\ref{Feynmanrules}), building up four triangles. 
The amplitude is the $\star$-product of the identification functions $\delta_{\minus x'_i}(x_i)$ of the boundary metric with the BF action for a single simplex $e^{i \sum_i \Tr x_i h_i}$,  $h_i\!=\!h_{t\tau} h_{\tau t'}$ being the parallel transport between the two triangles sharing $i$, integrated over the bulk connection $h_{t\tau}$. 
Using the above, one can see easily that the amplitude of generic open graphs is then given by a path integral for the BF action augmented by the appropriate boundary terms. This implies also that the BF action for a single simplex is already explicitly present in the interaction term of the GFT action. This can be useful to study the link with semi-classical/continuum gravity at the GFT level.


These results show an exact duality between spin foam models and simplicial gravity path integrals, stemming from two equivalent representations of the GFT field: as a function $\Phi^j_{mn}$ of representation labels, obtained by harmonic analysis, and as a non-commutative function $\widehat{\varphi}$ of Lie algebra variables, interpreted as metric variables.
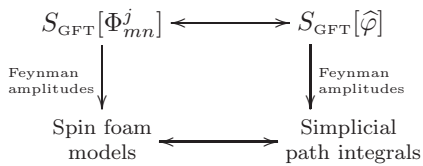
\[ 
\xymatrix{
  S_{\tiny \mbox{GFT}}[\Phi^j_{mn}] \ar@{<->}[r] \ar[d]_{\tiny \begin{matrix} \mbox{Feynman} \hfill \\ \mbox{amplitudes} \end{matrix} }& **[r]S_{\tiny \mbox{GFT}}[\widehat{\varphi}] \ar[d]^{\tiny \begin{matrix} \mbox{Feynman} \hfill \\ \mbox{amplitudes} \end{matrix} }
  \\
  {\footnotesize \begin{matrix} \mbox{Spin foam} \hfill\\ \mbox{models} \end{matrix}}   \ar@{<->}[r]& **[r]
  {\footnotesize \begin{matrix} \mbox{\,\, Simplicial} \hfill\\  \mbox{path integrals} \end{matrix} }
}
\]



{\bf Towards 4d gravity models.}
\noindent Going up dimensions, we consider the GFT model for $\SO(4)$ BF theory, defined in terms of a gauge invariant field $\vphi_{1234} \!=\! \int \extd h \, \vphi(hg_1, hg_2, hg_3, hg_4)$ by the action:
\beq
S =\! \frac{1}{2}\int \varphi^2_{1234} -  \frac{\lambda}{5!} \! \int \varphi_{1234}\,\varphi_{4567}\,\varphi_{7389}\,\varphi_{962\,10}\,\varphi_{10\,851}.
\eeq
The Feynman graphs are 2-complexes dual to 4d simplicial complexes: the combinatorics of the interaction term is that of a 4-simplex; the kinetic terms dictates the gluing rules for 4-simplices along tetrahedra. Using harmonic analysis on $\SO(4)$, the Feynman amplitudes take the form of the Ooguri state sum model.  


The $\SO(3)$ group Fourier transform naturally extends to a Fourier transform on $\SO(4)\simeq\SU(2) \times \SU(2) / \Z_2$, which is invertible on even functions 
$f(g)\!=\!f(\minus g)$. In the sequel we assume the further invariance of the Ooguri field under $g_i \to -g_i$ in each of the variables. 

The dual Ooguri field  is a function of four $\so(4)$ Lie algebra elements, or bivectors, associated to the four triangles of each tetrahedron. 
Gauge invariance translates into a closure constraint for the bivectors, meaning that the four triangles close to form a tetrahedron. Kinetic and vertex terms encode the identification, up to parallel transport, of the bivectors associated to the same triangle in different tetrahedral frames. As in 3d, Feynman amplitudes are simplicial path integrals for BF theory. 

The new representation of the Ooguri model provides a convenient starting point for imposing in a geometrically transparent manner the discrete simplicity constraint that turn BF theory into 4d simplicial gravity \cite{new}. Using the decomposition of $x\!\in\!\so(4)$ into selfdual $x^\plus$ and anti-selfdual $x^\minus$       $\su(2)$-components,  we impose that the four bivectors in each tetrahedron are orthogonal to the same vector $k \!\in\! \cS^3\!\sim\!\SU(2)$ normal to the tetrahedron, by means of the constraint projector 
\beq
\widehat{S}_k(x^\minus_j, x^\plus_j) =  \prod_{j=1}^4 \delta_{\minus kx^\minus_j k^\inv}(x^\plus_j)
\eeq
where the $\delta$ functions are given by (\ref{delta}). One can show that $\widehat{S}_k$ acts dually as the projector onto fields on the homogeneous space $\cS^3 \!\sim\! \SO(4)/ \SO(3)_k$, where $\SO(3)_k$ is the stabilizer of $k$. 
The case $k\!=\!1$ reproduces the standard Barrett-Crane projector. 

By combining the simplicity projector $\widehat{S}\!:=\!\widehat{S}_1$ with closure, one may build up the field  
$\widehat{\Psi}\!:=\!\widehat{S} \star \widehat{C}\star\vphihat$ of the standard GFT formulation of the Barrett-Crane model. More precisely,  combining the interaction term: 
\beq
\frac{\lambda}{5!}\int \widehat{\Psi}_{1234}\star\widehat{\Psi}_{4567}\star\widehat{\Psi}_{7389}\star\widehat{\Psi}_{962\,10}\star\widehat{\Psi}_{10\,851}
\eeq
with the possible kinetic terms:
\beq
\frac{1}{2}\int \widehat{\Psi}^{\star2}_{1234}\; , \hspace{0.1cm} \frac{1}{2} \int (\widehat{C}\star\vphihat)^{\star2}_{1234}\, \hspace{0.1cm} \text{or}\hspace{0.2cm} \frac{1}{2} \int \vphihat^{\star2}_{1234}
\eeq
gives the versions of the Barrett-Crane model derived in \cite{DPFKR}, \cite{PR} and \cite{valentin} respectively. 
The origin of these different versions can be understood geometrically, thanks to the new GFT representation.
Given $h\!\in\!\SO(4)$, one has
\beq
(e_h \star \widehat{S}_k)(x) 
= (\widehat{S}_{h \rhd k} \star e_h)(x)
\eeq
with $h \rhd k \!:=\!h^\plus k (h^\minus)^{\minus 1}$. This expresses the fact that, after rotation by $h$,  simple bivectors with respect to the normal $k$ become simple with respect to the rotated normal $h \rhd k$. Therefore closure and simplicity constraints do not commute. Moreover, whereas the model couples correctly the bivector variables $x$ across simplices, the integration over holonomies effectively decorrelates the normal vectors $k$ associated to the same tetrahedron in different 4-simplices. This implies a missing geometric condition on connection variables $h_{\tau\sigma}$. Work on a GFT model where simplicity constraints are imposed covariantly  is currently in progress.

A simplicial path integral formulation of the Barrett-Crane model, in, say, its version \cite{valentin}, is obtained by using the Feynman rules for the propagator and vertex:
\beq
\prod_{i=1}^4 (\delta_{\minus x_i})(y_i), 
\quad 
\int \prod_t \extd h_{\tau \sigma}  \prod_{i=1}^{10} ( \delta_{\minus x_i} \star \widehat{S} \star \e_{h_{\tau\tau'}})(y_i)
\eeq
The amplitude of a graph dual to a triangulation $\Delta$ takes the form of the $\star$-evaluation of a non-commutative observable in BF theory:
\beq
Z_{\mbox{\footnotesize BC}}(\Delta) = \int \prod_{\tau \sigma} \extd h_{\tau \sigma} \! \int \prod_t \extd^6 x_t \, (\cO_t \star \e_{H_t}) (x_t)
\eeq 
where the functions $\cO_t(x_t)$ implement simplicity $\delta_{\minus h_{0j}^{\minus \inv} x_t^{\minus} h_{0j}^{\minus}}(h_{0j}^{\plus \inv} x_t^\plus h_{0j}^{\plus})$ of the bivectors $x_t$ in each of the 4-simplex frames $j\!=\!0 \!\cdots\! N$ around $t$: 
\beq
\cO_t= \bigstar_{j=0}^N \, \delta_{\minus h_{0j}^{\minus \inv} x_j^\minus h_{0j}^{\minus}}(h_{0j}^{\plus \inv} x_j^\plus h_{0j}^{\plus}) \quad
\eeq


{\bf Conclusions and perspectives.}
\noindent The new non-commutative representation of GFTs introduced in this paper, based on the group Fourier transform, realizes an explicit GFT duality between spin foam models and simplicial gravity path integrals. It also makes explicit how simplicial geometry is encoded in the GFT formalism. 


Let us mention a few research directions, some already ongoing, stemming from our results. 

The interpretation of GFTs as a 2nd quantization of spin networks suggests to apply the group Fourier transform to generic loop quantum gravity states. This should give a {\sl flux} representation of the theory, usually assumed to be intractable precisely because of the non-commutativity of flux operators. 

The new representation should also help the identification of spacetime symmetries (e.g. diffeomorphisms \cite{diffeo}) which act on the B variables, at the level of the GFT action.  Understanding the role of diffeomorphisms can then guide the study of the relation between GFTs and continuum general relativity. 

Obviously, the goal is the construction of a satisfactory GFT model for quantum gravity in 4 dimensions. In the new GFT representation, guided by the manifest geometric meaning of variables and amplitudes, simplicity constraints on the B variables, with and without Immirzi parameter, can be imposed in a natural way. 
This is work in progress and can either lead to a new spin foam model for 4d quantum gravity, or to a geometrically clear GFT formulation of the recently proposed ones, in terms of simplicial path integrals.

The new representation may help also the study of GFT renormalization \cite{GFTren, linearized} and that of their phase structure and continuum approximation \cite{gftfluid}. It can also be used for the introduction of scales, re-expressing the star product in terms of differential operators \cite{NCFourier}.

Finally, it should reinforce the links between the GFT formalism and non-commutative geometry, as well as the approach to quantum gravity phenomenology \cite{QGphen} based on effective non-commutative matter field theories.


{\bf Acknowledgements.}
We thank B. Dittrich, F. Girelli and J. Tambornino for discussions. Support from the A. von Humboldt Stiftung
through a Sofja Kovalevskaja Prize is acknowledged.

\end{document}